\documentclass[11pt,a4paper,english]{article}

\usepackage{epsfig}
\usepackage{amsfonts}
\usepackage{amssymb,amsmath}
\usepackage[ps,matrix,arrow]{xy}
\usepackage{eucal}

\makeatletter

\catcode`@=11
\renewcommand{\section}
{\@startsection{section}{1}{0pt}{\medskipamount}{\medskipamount}{\large\bf}}
\makeatletter\renewcommand{\subsection}
{\@startsection{subsection}{2}{\z@}{-3.25ex plus -1ex minus -.2ex}
{1.5ex plus .2ex}{\it }}

\numberwithin{equation}{section} \catcode`@=12

\def\e{{\,\rm e}\,}

\def\ii{{\,{\rm i}\,}}

\def\beq{\begin{equation}}
\def\bee{\begin{equation}}
\def\eeq{\end{equation}}
\def\bea{\begin{eqnarray}}
\def\eea{\end{eqnarray}}
\def\bd{\begin{displaymath}}
\def\ed{\end{displaymath}}

\newcommand{\Cint}{\int\kern-10.5pt-\kern7pt}

\newcommand{\be}{\begin{equation}}
\newcommand{\ee}{\end{equation}}

\newcommand\fverb{\setbox\pippobox=\hbox\bgroup\verb}
\newcommand\fverbdo{\egroup\medskip\noindent%
                        \fbox{\unhbox\pippobox}\ }
\newcommand\fverbit{\egroup\item[\fbox{\unhbox\pippobox}]}
\newbox\pippobox

{

\def\be{\begin{equation}}
\def\ee{\end{equation}}
\def\bea{\begin{eqnarray}}
\def\eea{\end{eqnarray}}

\begin{document}

\begin{titlepage}
\setcounter{page}{0}
\begin{flushright}
ITP--UU--08/21 , SPIN--08--19\\
DAMTP--2008--26\\
HWM--08--2 , EMPG--08--4
\end{flushright}

\vskip 1.8cm

\begin{center}

{\Large\bf Instantons and Donaldson--Thomas Invariants}

\vspace{15mm}

{\large\bf Michele Cirafici$^{(a)}$, Annamaria Sinkovics$^{(b)}$
and
 Richard~J.~Szabo$^{(c)}$}
\\[6mm]
\noindent{\em $^{(a)}$ Institute for Theoretical Physics and
Spinoza
  Institute \\
Utrecht University, 3508 TD Utrecht, The Netherlands \\Email: {\tt
  M.Cirafici@uu.nl}}\\[3mm]
\noindent{\em $^{(b)}$ Department of Applied Mathematics and
  Theoretical Physics
\\ Centre for Mathematical Sciences, University of
Cambridge \\ Wilberforce Road, Cambridge CB3 0WA, UK\\ Email: {\tt
  A.Sinkovics@damtp.cam.ac.uk}}\\[3mm]
\noindent{\em $^{(c)}$ Department of Mathematics, Heriot--Watt
  University and\\
Maxwell Institute for Mathematical Sciences\\
Colin Maclaurin Building, Riccarton, Edinburgh EH14 4AS, UK\\
Email: {\tt
  R.J.Szabo@ma.hw.ac.uk}}

\vspace{15mm}

\begin{abstract}
\noindent

We review some recent progress in understanding the relation
between a six dimensional topological Yang--Mills theory and the
enumerative geometry of Calabi--Yau threefolds. The gauge theory
localizes on generalized instanton solutions and is conjecturally
equivalent to Donaldson--Thomas theory. We evaluate the partition
function of the $U(N)$ theory in its Coulomb branch on flat space
by employing equivariant localization techniques on its
noncommutative deformation. Geometrically this corresponds to a
higher dimensional generalization of the ADHM formalism. This
formalism can be extended to a generic toric Calabi--Yau.

\end{abstract}

\end{center}

\vspace{1.5cm}

\noindent {\em Contribution to the proceedings of the 3rd workshop
of the RTN project `Constituents, Fundamental Forces and
Symmetries of the Universe' in Valencia, 1-5 October, 2007.}

\end{titlepage}


\newpage

\allowdisplaybreaks

\section{Introduction}

Topological field and string theories have been the focus of
extensive investigation in the last two decades. These models are
more tractable than their physical counterparts but still capture
some interesting physical quantities, in particular those related
to the vacuum structure of the full quantum theory. Due to the
topological nature of the model these quantities can be often
computed exactly. The underlying reason is their deep relation to
geometric and topological invariants of the physical space where
the original model is defined.

In this note we will focus on the Donaldson--Thomas (DT)
invariants \cite{Donaldson:1996kp}. From the viewpoint of the
topological string they can be defined as follows. One starts with
a Calabi--Yau manifold $X$ on which the topological A--model is
defined. Then the DT invariant corresponds to the number of bound
states formed by a single D6 brane wrapping the full Calabi--Yau
manifold with a D2 brane wrapping a 2--cycle $C \subset X$ in
homology class $\beta$ and $m$ D0 branes. This configuration is
encoded in a mathematical object called an \textit{ideal sheaf}
and the set of all possible configurations is described by the
moduli space of ideal sheaves $I_{m} (X,\beta)$. This space is
also known as the Hilbert scheme of points and curves of the
threefold $\mathrm{Hilb}^m (X , \beta)$. Then the DT invariant
$D^m_{\beta} (X)$ is defined as the ``volume'' of this moduli
space.

If the Calabi--Yau is toric all the geometric information can be
essentially encoded in a combinatorial problem and the topological
string has a reformulation in terms of the classical statistical
mechanics of a melting crystal \cite{Okounkov:2003sp}. In this
more physical setting the DT invariants parametrize the atomic
configurations of the melting crystal. This leads to a very
non--trivial conjecture that the geometrical information captured
by the DT invariants is equivalent to Gromov--Witten theory, since
they are two different expansions of the same topological string
amplitude. So far this conjecture has been proven in a number of
cases \cite{MNOP}.

A detailed understanding of DT theory on Calabi--Yaus could
sharpen our knowledge about the geometrical meaning of the
topological string and thus about the vacuum structure of the full
string theory. In this note we will report about some progress
towards this ambitious goal \cite{Cirafici:2008sn}. Namely we will
only consider local toric threefolds on which the DT problem can
be conjecturally rephrased as a topological gauge theory
\cite{Iqbal:2003ds}. We will put this conjecture on firmer grounds
and by employing the techniques of equivariant localization show
how to set the ground for explicit computations. We will apply our
formalism to higher rank DT invariants on the Coulomb branch of
the gauge theory.

\section{The Topological Gauge Theory and Equivariant Localization}

Let us consider a local toric threefold $X$. In this case DT
theory can be (conjecturally) described by a six dimensional
abelian topological gauge theory living on the worldvolume of the
D6 brane wrapping $X$ \cite{Iqbal:2003ds}. This gauge theory is
the topologically twisted version of maximally supersymmetric
Yang--Mills in six dimensions
\cite{Blau:1997pp}--\cite{Acharya:1997gp}. Its bosonic matter
content consists of a gauge field $A_{\mu}$, a complex Higgs field
$\Phi$ and a $(3,0)$ form $\rho^{3,0}$ along with their complex
conjugates. Essentially its action has the form of the topological
density
\begin{equation} \label{instaction}
\frac{1}{2}\, \int_X \, \mathrm{Tr} \Big(   F_A \wedge F_A \wedge
k_0 + \mbox{$\frac\vartheta3$}\, F_A \wedge F_A \wedge F_A \Big) \
,
\end{equation}
supplemented by a gauge fixing term. Here $k_0$ is the background
K{\"a}hler two-form of $X$ and $\vartheta$ is the six-dimensional
theta-angle which will be identified with the topological string
coupling~$g_s$. This gauge theory has a BRST symmetry and hence
localizes onto the moduli space of solutions of the fixed point
equations
\begin{eqnarray} \label{inste} F_A^{2,0} &=&
  \overline{\partial}\,_A^{\dagger} \rho
  \ , \nonumber\\[4pt]
F_A^{1,1} \wedge k_0 \wedge k_0 + [\rho, \overline{\rho}\,] &=&
l~k_0 \wedge k_0
\wedge k_0 \ , \nonumber\\[4pt] \mathrm{d}_A \Phi &=& 0 \ .
\end{eqnarray}
The solutions of these equations minimize the gauge theory action
and we will therefore call them generalized instantons or just
instantons. On a Calabi--Yau manifold we can set the field $\rho$
to zero. Then the first two equations reduce to the
Donaldson--Uhlenbeck--Yau (DUY) equations which are conditions of
stability for holomorphic bundles over~$X$ with finite
characteristic classes.

The introduction of this auxiliary gauge theory essentially
reformulates DT theory as a (generalized) instanton counting
problem. The gauge theory localizes onto the moduli space
$\mathcal{M} (X)$ of holomorphic bundles (or coherent sheaves) on
$X$ and the instanton multeplicities in the instanton expansion of
the path integral represent the DT invariants. Note that in the
gauge theory language it is immediate to generalize DT theory to a
non--abelian $U(N)$ setting, with an arbitrary number of D6 branes
(corresponding to generic rank $N$ bundles).

So far we have reduced the difficult algebro--geometrical problem
of counting sheaves to a more tractable path integral.
Unfortunately the theory as it stands is not very manageable since
moduli spaces of instantons suffer from non-compactness problems
arising both from singularities where instantons shrink to zero
size as well from the non-compactness of the ambient space $X$ on
which the gauge theory is defined.
 The way out comes from an
analogous issue in instanton counting in four dimensional twisted
$\mathcal{N}=2$ theory. In \cite{Nekrasov:2002qd} Nekrasov
proposed that equivariant localization techniques could be used in
combination with a noncommutative deformation of the theory to
evaluate directly the instanton factors. This idea has turned out
to be very powerful allowing for explicit computations in the four
dimensional setting and can be applied to our six dimensional case
\cite{Iqbal:2003ds}. Here we will only consider the case of flat
space $X = \mathbf{C}^3$ unless explicitly mentioned. The
noncommutative deformation resolves the small instanton
singularities of the moduli space $\mathcal{M} (X)$ and provides a
natural compactification of $\mathcal{M} (X)$. Also working
equivariantly can be easily implemented: on $\mathbf{C}^3$ there
is naturally the action of the torus $\mathbf{T}^3$ coming from
the maximal torus of the $U(3)$ group generating rotational
isometries that preserve the K\"ahler form of $\mathbf{C}^3$. On
the coordinates of $\mathbf{C}^3$ this torus acts as $z_i
\longrightarrow z_i \e^{\ii \epsilon_i}$, $i=1,2,3$.

The equivariant model can be obtained by modifying the BRST
operator so that it becomes an equivariant differential with
respect to this toric action. In other words we restrict attention
to field configurations that are annihilated by the old BRST
operator only up to a toric action.

After these modifications the gauge theory localizes onto the
fixed points of the equivariant BRST operator. One can show that
these fixed points are isolated and their contribution to the path
integral can be computed by direct equivariant integration, by
using the Duistermaat--Heckman formula or its generalizations. The
problem of computing the path integral is now reduced to two
simpler ones, namely the classification of the critical points of
the equivariant BRST differential and the actual evaluation of the
instanton factor.

These goals can be accomplished in two distinct but ultimately
equivalent ways as we are about to see.

\section{The Noncommutative Theory}

The path integral of the noncommutative field theory can be
evaluated directly by using equivariant localization. After the
noncommutative deformation we can think of the theory as an
infinite--dimensional matrix model where the fields are replaced
by operators acting on a separable Hilbert space. This approach
has the advantage that some explicit instanton solutions can be
constructed and it provides a natural compactification of the
instanton moduli space. In terms of the noncommutative fields the
instanton equations become
\begin{eqnarray}
\big[Z^{i}\,,\, Z^{j}\big] + \epsilon^{i
j k} \,\big[Z^{\dagger}_{k}\,,\, \rho\big] &=& 0 \ , \nonumber\\[4pt]
\big[Z^{i}\,,\, Z^{\dagger}_{i}\big] + \big[\rho\,,\,
\rho^{\dagger}
\big] &=& 3~1_{N\times N} \ , \nonumber\\[4pt]
\big[Z_{i} \,,\, \Phi\big] &=& \epsilon_{i}\, Z_{i} \ ,
\label{adhmform}
\end{eqnarray}
where in the last equation there is no sum over the index $i$ and
the right--hand side reflects explicitly the equivariant
deformation.

These sets of equations can be solved by three-dimensional
harmonic oscillator algebra. The unique irreducible representation
of this algebra is provided by the Fock module
\begin{equation}
\mathcal{H}=\mathbf{C}
\big[\alpha_1^\dagger\,,\,\alpha_2^\dagger\,,\,
\alpha_3^\dagger\big]|0,0,0\rangle=\bigoplus_{n_1,n_2,n_3\in
\mathbf{N}_0}\, \mathbf{C} |n_1,n_2,n_3\rangle \ ,
\label{Fockspdef} \end{equation} where $|0,0,0\rangle$ is the Fock
vacuum with $\alpha_i|0,0,0\rangle=0$ for $i=1,2,3$, and the
orthonormal basis states $|n_1,n_2,n_3\rangle$ are connected by
the usual action of the creation and annihilation operators
$\alpha_i^\dagger$ and $\alpha_i$. The operators $Z^i$ may then be
taken to act on the Hilbert space $ \mathcal{H}_W =W\otimes
\mathcal{H} $ where $W\cong\mathbf{C}^N$ is a Chan-Paton
multiplicity space of dimension $N$, the number of D6-branes (and
the rank of the gauge theory). The space $W$ carries the
nonabelian degrees of freedom and we understand $Z^i$ and $\Phi$
as $N \times N$ matrices of operators acting on $\mathcal{H}$.

We can diagonalize the field $\Phi$ using the $U(N)$ gauge
symmetry. One can now classify the fixed points of the nonabelian
gauge theory by generalizing the arguments
of~\cite{Iqbal:2003ds,swpart}. We are prescribed to compute the
path integral over configurations of the Higgs field whose
asymptotic limit is $\mathbf{a} = \mathrm{diag}(a_1 , \dots ,
a_N)\in u(1)^N$. With this choice of boundary condition the
noncommutative field $\Phi$ has the form $ \Phi = \mathbf{a}
\otimes {1}_{\mathcal{H}} + {1}_{N\times N} \otimes
\Phi_{\mathcal{H}} $. The degeneracies of the asymptotic Higgs
vevs breaks the gauge group $U(N)\to\prod_i\,U(k_i)$ with
$\sum_i\,k_i=N \ . $ Correspondingly, the Chan-Paton multiplicity
space $W$ decomposes into irreducible representations
$W=\bigoplus_i\,W_i$ with $\dim_\mathbf{C} W_i=k_i$.

Due to the equivariant deformation the theory now localizes on
$U(1)^N$ noncommutative instantons. These correspond to ideals
$\mathcal{I}$ of codimension $k$ in $\mathbf{C} [z_1 , z_2 , z_3]$
that are associated, via partial isometries, to subspaces of the
full Hilbert space of the form $\bigoplus_{f\in\mathcal{I}}\,
f\big(\alpha_1^\dag\,,\,\alpha_2^\dag\,,\,\alpha_3^\dag
\big)|0,0,0\rangle \ $. These ideals are generated by monomials
$z^i\,z^j\,z^k$ and are in one-to-one correspondence with
three-dimensional partitions, with the triplet $(i,j,k)$
corresponding to boxes of the partition. More precisely, the set
of solutions can be completely classified in terms of coloured
partitions $ \vec\pi=(\pi_1, \ldots, \pi_N)$, which are rows of
$N$ ordinary three-dimensional partitions $\pi_l$ labelled by
$a_l$.

We can now write the full path integral as a sum over critical
points and compute the fluctuation factor around each critical
point. This factor assumes the form of a ratio of functional
determinants
\begin{equation}
\frac{\det \left(\mathrm{ad}\, \Phi \right)\, \det
\left(\mathrm{ad}\, \Phi + \epsilon_1 + \epsilon_2 \right)\, \det
\left(\mathrm{ad}\, \Phi + \epsilon_1 + \epsilon_3 \right)\, \det
\left(\mathrm{ad}\, \Phi + \epsilon_2 + \epsilon_3 \right)}{\det
\left(\mathrm{ad}\, \Phi+ \epsilon_1 + \epsilon_2 + \epsilon_3
\right)\, \det \left(\mathrm{ad}\, \Phi + \epsilon_1 \right)\,
\det \left(\mathrm{ad}\, \Phi + \epsilon_2 \right)\, \det
\left(\mathrm{ad}\, \Phi + \epsilon_3 \right)} \ ,
\end{equation}
where the $\epsilon_i$ parametrize the toric action. This ratio
can be computed explicitly to give a factor of $(-1)^{N
|\vec\pi|}$. Combined with a similar computation for the instanton
action, this gives the instanton expansion
\begin{equation} \label{Znc}
\mathcal{Z}_{\rm DT}^{U(1)^N}\big(\mathbf{C}^3\big)
=\sum_{\vec\pi}\, (-1)^{(N+1)\,|\vec\pi|}~q^{|\vec\pi|} \ ,
\end{equation}
where $q = - \e^{\ii \vartheta}=\e^{-g_s}$.

\section{Matrix Quantum Mechanics and Coherent Sheaves}

The second approach consists in the introduction of an appropriate
topological matrix quantum mechanics \cite{Cirafici:2008sn}. From
the string theoretical point of view this corresponds to the
effective action on a gas of D0 branes that bound the original D6
on $\mathbf{C}^3$. From the perspective of the gauge theory it
arises as quantization of the collective coordinates around each
instanton solution and provides a higher dimensional
generalization of the ADHM construction of instantons. Indeed one
can see this explicitly by parametrizing each holomorphic bundle
(or coherent sheaf) on the projective space $\mathbf{P}^3$ that
corresponds to a compactification of the physical space
$\mathbf{C}^3$ in term of a set of algebraic matrix equations that
we'll call generalized ADHM equations. This can be done by using
Beilinson's theorem which states that for any coherent sheaf
$\mathcal{E}$ on $\mathbf P^3$ there is a spectral sequence
$E_s^{p,q}$ with $E_1$-term $E_1^{p,q} = H^q \big( {\mathbf P}^3
\,,\, \mathcal{E}(-r) \otimes \Omega^{-p}_{{\mathbf P}^3}(-p)
\big) \otimes \mathcal{O}_{{\mathbf P}^3}(p)$ for $p \le 0$ that
converges to the original sheaf (here $\Omega_{\mathbf{P}^3}$ and
$\mathcal{O}_{\mathbf{P}^3}$ are respectively the sheaf of
differential forms and the structure sheaf). By the appropriate
set of boundary conditions this spectral sequence degenerates at
the $E_2$ term.

The outcome of this procedure is that the original sheaf can be
described as the only non--vanishing cohomology of a four term
complex. The associated conditions yield a particular set of
matrix equations plus stability conditions. One can show that this
system boils down to the following set of generalized ADHM
equations
\begin{equation}
[B_1,B_2]+I\,J = 0 \ , \qquad [B_1,B_3]+I\,K = 0 \ , \qquad
[B_2,B_3] = 0 \ , \label{ADHMeqs}\end{equation} where  $B_i\in
\mathrm{End}(V)$, $i=1,2,3$, $I\in\mathrm{Hom}(W,V)$ and
$J,K\in\mathrm{Hom}(V,W)$ and a suitable stability condition has
to be imposed. The vector spaces $V$ and $W$ arise in the
geometrical construction outlined above as particular cohomology
groups of the sheaf $\mathcal{E}$. These equations are naturally
in correspondence with the noncommutative instantons described in
the previous section. In particular in the abelian case $N=1$ the
stability conditions allow us to set $J=K=0$ and the cohomology
sheaf $\mathcal{E}$ is isomorphic to the ideal $\mathcal{I}$ that
enters in the description of the Hilbert scheme in terms of
noncommutative instantons. As we localize the theory onto its
$U(1)^N$ phase this is the relevant case. One can easily construct
a cohomological matrix model starting from these equations. In
this framework $V$ with $\mathrm{dim} V = k$ represent the gas of
$k$ D0 branes (or the charge $k$ topological sector in the gauge
theory) while $W$ stands for the D6 branes and its dimension is
the rank of the gauge theory $N$.

The matrices $B_i$ arise from 0--0 strings and represent the
position of the coincident D0-branes inside the D6-branes. On the
other hand, the field $I$ describes open strings stretching from
the D6-branes to the D0-branes. It characterizes the size and
orientation of the D0-branes inside the D6-branes. Other fields
are necessary to close the equivariant BRST algebra and localize
the theory on the generalized ADHM equations but we refer the
reader to \cite{Cirafici:2008sn} for a complete treatment.

In the abelian case the generalized ADHM equations ensure that the
critical points can be expressed by a certain sequence of maps
between the spaces $V$ and $W$. This configuration can be
explicitly mapped into a three dimensional partition thus
recovering the classification of the fixed points that we found in
the noncommutative setting. The generalization to the $U(1)^N$
theory is simple and corresponds to $N$--tuples of three
dimensional partitions. We will denote a generic fixed point $f$
as $\vec{\pi} = (\pi_1 , \dots , \pi_N)$. Accordingly at the fixed
points the vector spaces $V$ and $W$ have the following weight
decompositions
\begin{equation} \label{decompos}
V_f = \sum_{l=1}^N \e^{ \ii a_l} \sum_{(i,j,k)\in \pi_l} t_1^{i-1}
t_2^{j-1} t_3^{k-1} \ , \qquad W_f = \sum_{l=1}^N \e^{ \ii a_l} \
,
\end{equation}
where $t_i = \e^{ \ii \epsilon_i}$ generate the $\mathbf{T}^3$
action.

The computation of the instanton factors proceeds as in the four
dimensional case \cite{Nekrasov:2002qd,Bruzzo:2002xf}. For every
fixed point we can describe the local structure of the moduli
space via an equivariant complex that encodes the linearization of
the generalized ADHM equations up to linearized (complexified)
gauge invariance. The character of this complex
\begin{equation} \label{character} \chi_{f} (\mathbf{C}^3)^{[k]} =
W_f^* \otimes V_f - \frac{ V_f^* \otimes W_f}{t_1 t_2 t_3} + V_f^*
\otimes V_f \frac{(1-t_1) (1-t_2) (1-t_3)}{t_1 t_2 t_3} \ ,
\end{equation}
contains all the local information needed in the localization
formula and can be used to compute explicitly the instanton
factors following \cite{Nekrasov:2002qd,Bruzzo:2002xf}. In
(\ref{character}) the subscript $f$ is stressing that the
computation only holds at a particular fixed point and the
conjugation acts on the elements of the weight decomposition as
$t_i^* = t_i^{-1}$. A straightforward computation gives the
fluctuation factor $(-1)^{N |\vec \pi|}$. To get the partition
function the last missing ingredient is the instanton action
(\ref{instaction}). This can be obtained by writing the universal
sheaf $\mathcal{E}$ on the moduli space as $\mathcal{E} = W \oplus
V \otimes (S^- \ominus S^+ )$ where $S^{\pm}$ are the
positive/negative chirality spinor bundles over $\mathbf{P}^3$. By
using the correspondence between spinors and differential forms we
can decompose its Chern character at a given fixed point as
\begin{equation}
\mathrm{ch} (\mathcal{E}_{\vec \pi}) = W_{\vec \pi} - (1-t_1)
(1-t_2) (1-t_3) V_{\vec \pi} \, .
\end{equation}
Collecting all pieces of information one can write down the full
partition function \begin{equation} \mathcal{Z}_{\rm
DT}^{U(1)^N}({\mathbf C}^3) =\sum_{f = \{ \pi_1 \cdots \pi_N \} }
(-1)^{N \sum_{l=1}^N |\pi_l|} {\, {\rm e} \,} ^{{\, {\rm i} \,}
\vartheta \sum_{l=1}^N |\pi_l|}  \ ,
\end{equation} that agrees precisely with (\ref{Znc}).

\section{The Coulomb Branch on a Toric Manifold}

The construction just outlined carries on to the case of a general
toric manifold $X$. The geometric information of a toric manifold
is encoded in a trivalent graph $\Delta$. The vertices $f$ of
$\Delta$ are locally isomorphic to $\mathbf{C}^3$ while the edges
$e$ corresponds to rational curves of K\"ahler area $t_e$. By
using the localization procedure on the physical space the gauge
theory localizes onto a sum of contributions associated with each
vertex corresponding to three dimensional partitions and a set of
propagators associated with the edges. Each propagator depends on
the area $t_e$ of the rational curve and on a two dimensional
partition that arises when gluing together two different three
dimensional partitions as a section of the common leg. In the rank
1 case one recovers the Calabi--Yau crystal picture directly from
the gauge theory. In the more general rank $N$ setting on the
Coulomb branch one finds
\begin{equation}
\mathcal{Z}_{\rm DT}^{U(1)^N} (X) = \sum_{\vec\pi_f}
q^{\mathbf{I}} (-1)^{(N+1) \mathbf{I}} \prod_{e \in \mathrm{edge}}
(-1)^{\sum_{l,l'=1}^N |\lambda_{l,e}| |\lambda_{l',e}| m_{1,e}}
{\, {\rm e} \,}^{- \sum_{l=1}^N |\lambda_{l,e}| t_e} \ ,
\end{equation}
where $q= -{\,\rm e}\,^{ {\,{\rm i}\,} \vartheta}$ and
\begin{equation}
\mathbf{I} = \sum_f \sum_{l=1}^N |\pi_{l,f}| + \sum_{e \in
\mathrm{edge}} \sum_{l=1}^N \sum_{(i,j) \in \lambda_{e,l}} \left(
m_{1,e} (i-1) + m_{2,e} (j-1) + 1 \right) \ .
\end{equation}
The integers $m_{1,e}$ and $m_{2,e}$ determine the normal bundle
to the rational curve corresponding to the edges.

\section{Conclusions}

In \cite{Cirafici:2008sn} we have studied the relationship with a
six dimensional topological Yang--Mills theory and
Donaldson--Thomas invariants. As a first step one can use
equivariant localization to write the partition function of the
noncommutative deformation of the theory as a sum over point--like
instantons. These noncommutative instantons can be interpreted in
purely geometrical terms as certain coherent sheaves on
$\mathbf{P}^3$ through a higher dimensional generalization of the
ADHM formalism. In turn this can be used to construct a
topological matrix quantum mechanics that dynamically describes
the stable coherent sheaves. This formalism can be used, for
example, to compute the rank $N$ partition function on a toric
manifold; the result is the $N$-th power of the abelian result
with an $N$ dependent sign shift. This shift can be absorbed in a
redefinition of the string coupling constant $g_s \longrightarrow
g_s - N {\rm i} \pi$. This modification is natural from the point
of view of the OSV conjecture \cite{Ooguri:2004zv} that relates
the entropy of a BPS black hole with the topological string
amplitude. The parameters that enter in the topological string
amplitude are functions of the D--brane charges at the attractor
point of the BPS moduli space. In the presence of D6 branes this
relation is consistent with the above shift.

\bigskip

\section*{Acknowledgments}

  M.C. wishes to thank the organizers of the workshop for
  providing such a stimulating atmosphere in such a beautiful
  location. This work was supported in part by the
Marie Curie Research Training Network Grants {\sl Constituents,
  Fundamental Forces and Symmetries of the Universe}
(Project~MRTN-CT-2004-005104) and {\sl Superstring Theory}
(Project~MRTN-CT-2004-512194) from the European Community's Sixth
Framework Programme, and by the INTAS Grant {\sl Strings, Branes
and
  Higher Spin Fields} (Contract~03-51-6346).

\end{document}